\documentclass[a4paper,10.5pt]{article}
\usepackage[tmargin=1in,bmargin=1in,lmargin=1in,rmargin=1in]{geometry}
\usepackage{url}
\usepackage[dvipsnames]{xcolor}
\usepackage[colorlinks = true,citecolor  = MidnightBlue,linkcolor  = NavyBlue,urlcolor   = MidnightBlue]{hyperref}
\usepackage{amsmath,amssymb,amsfonts}
\usepackage{natbib}
\usepackage{graphicx}
\usepackage{colortbl,booktabs}
\usepackage{setspace}
\newcommand{\R}{\mathbb{R}}
\newcommand{\p}{\mathbb{P}}
\newcommand{\bu}{{\bf u}}
\newcommand{\bU}{{\bf U}}
\newcommand{\cU}{{ {\mathcal U}}}
\newcommand{\bcU}{{\boldsymbol {\cU}}}
\newcommand{\bgamma}{{\boldsymbol \gamma}}
\newcommand{\btheta}{{\boldsymbol \theta}}
\newcommand{\bTheta}{{\boldsymbol \Theta}}
\newcommand{\bmu}{{\boldsymbol \mu}}
\newcommand{\CG}{\mathcal{G}}

\usepackage{dsfont}

\DeclareMathOperator{\argmin}{argmin}

\begin{document}

% Title of paper
\title{\LARGE \textbf{Structure estimation of binary graphical models on stratified data: application to the description of injury tables for victims of road accidents}}

\author{\scshape{Nadim BALLOUT and Vivian VIALLON}}

\date{\textit{\small Univ. Lyon, Universit\'e Claude Bernard Lyon 1, Ifsttar, UMRESTTE, UMR T$\_$9405, F-69373 LYON.}\\ 
	\text{\small nadim.ballout@ifsttar.fr}\\	}
\maketitle

\begin{abstract}
 Graphical models are used in many applications such as medical diagnostic, computer security, etc. More and more often, the estimation of such models has to be performed on several predefined strata of the whole population. For instance,  in epidemiology and clinical research, strata are often defined according to age, gender, treatment or disease type, etc. In this article, we propose new approaches aimed at estimating binary graphical models on such strata. Our approaches are obtained by combining well-known methods when estimating one single binary graphical model, with penalties encouraging structured sparsity, and which have recently been shown appropriate when dealing with stratified data. Empirical comparions on synthetic data highlight that our approaches generally outperform the competitors we considered. An application is provided where we study associations among injuries suffered by victims of road accidents according to road user type.\vspace{0.3cm}
\end{abstract}

\section{Introduction}
In this article, we consider the estimation of the conditional dependence structure among a set of binary variables across several predefined sub-groups, or strata, of a population. As an illustration, we will consider the description of the injury tables suffered by victims of road accidents. This description is key to the quantification of the needs in terms of care services and thus, in a long term perspective, to improve the care of the victims. Fine description of the associations among injuries could further turn out to be useful for diagnostic purposes: if a given external, and so easy to diagnose, injury is strongly positively associated with some internal and harder to diagnose injury, then special attention would be given to a victim suffering from the external injury as this victim is more likely to suffer from the internal one as well. For all these reasons, clinicians ask for statistical tools that can accurately summarize injury tables of victims, as well as the associations among injuries. Of course, the characteristics of the accident play an important role in the set of injuries suffered by the victims. In particular, injury tables are likely to vary according to the road user type of the victim: pedestrian, car occupant, motorized two-wheeler user, cyclist, etc. Therefore, associations among injuries have to be studied according to road user types, that is, across several predefined strata of the population of victims. \\
Graphical models have been recognized as valuable tools to model the joint distribution of $p$ given variables as well as graphically representing the conditional dependences between them \citep*{lauritzen1996}. A graphical model is a non-directed graph that consists of a set of $p$ nodes corresponding to the $p$ variables, along with a set of edges joining some pairs of nodes. More precisely, an edge is absent between two nodes if and only if the two corresponding variables are conditionally independent given the other variables. Regarding the application we have in mind, each injury can be modeled by a binary random variable that equals $1$ if the victim suffers from this injury and $0$ otherwise. The description of the injury tables of victims then reduces to the description of the joint distribution of $p$ binary variables, where $p$ is the number of all possible injuries. When working with binary variables, the quadratic exponential binary model, also known as the Ising model, is commonly used \citep*{besag1974, cox1994}. Identifying the structure of the binary graphical model reduces to the determination of non-null parameters in the Ising model (see Section \ref{sec:Ising} below for details), and then to a model selection problem. However, when $p$ is larger than 20 or 30 variables,  inference under the Ising model is difficult because of the intractability of the log-partition function. In particular, maximum likelihood estimation can generally not be performed. Various solutions have arisen in the literature. \citet*{wainwright2007} proposed to use multiple $\ell_1$-penalized logistic regressions, extending the approach developed by \citet*{meinshausen2006} in the Gaussian case. Following the terminology adopted in \citet*{wang2010}, we will refer to this approach as SepLogit. \citet*{hofling2009} considered a variant based on $\ell_1$-penalized pseudo-likelihood. \citet*{banerjee2008} derived a Gaussian approximation of the Ising log-likelihood, while \citet*{yang2011} used variational inference based on alternative approximations of the log-partition function. These approaches have been empirically compared in \citet{viallon2014}. Under the designs they considered, all these methods performed similarly, and reasonably well, to identify the structure of one binary graphical model on one single population. \\
When models have to be estimated on several predefined strata of a population, the general objective is to take advantage of the potential homogeneity among the corresponding models, while not masking the heterogeneities. Several authors have studied the estimation of regression models on such stratified data. \citet{ViallonFused} as well as \citet*{tutz2010, tutz2012} studied generalized fused lasso estimates under generalized linear models. \citet*{gross2016} and \citet*{ollier2017} independently developed an alternative approach referred to as data shared lasso. It can be seen as an extension of a common strategy which consists in first selecting a reference stratum and then adding interaction terms between each covariate and the indicators of the remaining strata. By considering an appropriate overparametrization, data shared lasso bypasses the arbitrary choice of the reference stratum and mimics the strategy based on an optimal and covariate-specific choice of the reference stratum. \citet*{ollier2017} established the sparsistency of data shared lasso under some technical assumptions in the case of linear regression models. In particular, for each covariate, data shared lasso is able to identify the strata on which the effects differ from those on the optimal reference stratum under nearly the same assumptions as those required by the optimal (and infeasible in practice) strategy. From a practical perspective, data shared lasso can be written as a weighted lasso on a simple transformation of the original data. It is therefore easy to implement under a variety of regression models, since very efficient lasso solvers are now available under many regression models: for instance the glmnet package is now available in R, Matlab and Python and uses several tricks to make the implementation extremely fast, such as strong rules for discarding predictors (\citet{tibshirani2012}; \citet*{elghaoui2012}).\\
As for the estimation of graphical models on stratified data,   \citet*{danaher2014} developed an approach based on a fused lasso penalty under Gaussian graphical models. For binary graphical models, \citet{Guo} based their approach on the pseudo-likelihood and a multiplicative decomposition of the coefficients. However, as will be made clearer below, this approach is not tailored to identify the heterogeneities that may exist between the conditional dependence structures of the different strata. In this article, we propose to combine the SepLogit method with either the generalized fused lasso or data shared lasso to jointly estimate binary graphical models on several predefined strata of a population. In Section \ref{sec:Methods}, we briefly recall some basics about the Ising model and the SepLogit method. Then, we describe our proposals and explain why they are better suited to identify heterogeneities than Guo and other's approach. In Section \ref{sec:Simul}, we present results from an empirical study, which establishes that our approaches outperform its competitors under the settings we consider. It further shows that our two proposals perform similarly. Section \ref{sec:Registre} presents the application of our approaches to describe injury tables on a French registry of victims of road accidents. Possible extensions are discussed in Section \ref{sec:Disc}. 

\section{Methods}\label{sec:Methods}
	\subsection{The Ising Model}\label{sec:Ising}
	The injury table of a victim can be modeled as a realization of the random variable
	$\bU=(U_1,...,U_p)^T \in \{0,1\}^p$ where $U_j$ indicates the presence of the injury $j$ in the considered injury table and $p$ is the cardinality of the set of all possible injuries. The description of the injury tables then reduces to the estimation of the joint distribution of  $\bU$, given an $n$-sample $\bU_1,..., \bU_n$ of independent and identically distributed (i.i.d.) replicas of $\bU$. The Ising model assumes the existence of a parameter vector $\bTheta^*=((\theta^*_{j})_{1\leq j\leq p}, (\theta^*_{j,\ell})_{1\leq j<\ell\leq p})^T$ in $\R^{p(p+1)/2}$ such that for any vector
	$\bu=(u_1,...,u_p) \in \{0,1\}^p$, the probability of the event $\{\bU = \bu\}$ is given by
	\begin{equation}\label{Ising}
		\p_{\bTheta^*}(\bU = \bu) = \exp\Big\{\sum_{j=1}^p\theta^*_j u_j+
		\sum_{j=1}^{p-1}\sum_{\ell=j+1}^p\theta^*_{j,\ell}u_ju_\ell
		- A(\bTheta^*)\Big\}.
	\end{equation}
	The so-called $\log$ \emph{partition function} $A: \R^p\rightarrow \R$,  is a normalization term ensuring that $\sum_{\bu\in\{0,1\}^p} \p_{\bTheta}(\bU =\bu)=1$ for every $\bTheta\in\R^{p(p+1)/2}$, and is defined as
	\begin{equation}\label{logpart}
		A(\bTheta) = \log
		\sum_{\bu\in\{0,1\}^p}\exp\left(\sum_{j=1}^p\theta_ju_j
		+\sum_{j=1}^{p-1}\sum_{\ell=j+1}^p\theta_{j,\ell}u_ju_\ell\right).
	\end{equation}		
	For every $j>\ell$, let $\theta^*_{j,\ell}=\theta^*_{\ell,j}$. For every $\bu=(u_1,...,u_p)\in\{0,1\}^p$ and every $j\in[p]$, where $[p]$ is the set of values $\{1,...,p\}$, further denote by $\bu_{-j}=(u_1,..., u_{j-1},u_{j+1},...,u_p)^T\in\{0,1\}^{p-1}$ the vector obtained  after the elimination of the $j$th component of $\bu$. Under (\ref{Ising}), we have, for every $j\in[p]$,
	\begin{equation}\label{bingraph_logist}
		\mbox{logit} \{\p_{\bTheta^*}(U_j=1|\bU_{-j} = \bu_{-j})\}=\ \theta^*_j + \sum_{\ell\neq
			j}\theta^*_{j,\ell}u_\ell = \theta^*_j + \btheta^{*T}_{-j} \bu_{-j},
	\end{equation}
	with $\btheta^*_{-j} = (\theta^*_{j,1},..,\theta^*_{j,j-1},\theta^*_{j,j+1},...,\theta^*_{j,p})^T$. Therefore, parameters $\theta^*_{j,\ell}$ correspond to the conditional log odds-ratios and conditional independence between $U_j$ and $U_\ell$ is equivalent to $\theta^*_{j,\ell}=0$. The Ising model is naturally associated to a graphical model, that is a non-directed graph $\CG =
	(V,E)$. The $p$ vertices of set $V$ correspond to the $p$ components of $\bU$. The set of edges $E\subseteq\{(j,\ell)\in V^2: j< \ell\}$ describes the conditional independence relationships among these components. More precisely, the edge $(j,\ell)$ between the nodes $j$ and $\ell$ is absent if and only if $U_j$ and $U_\ell$ are independent conditionally on the other variables, that is if and only if $\theta^*_ {j,\ell}=0$. Therefore, the identification of the edge set, or the structure of the graph, reduces to the identification of the zeros in the vector $\bTheta^*$. However, the
	estimation of $\bTheta^*$ and the selection of the non-zero elements of $\bTheta^*$ under the Ising model is not straightforward 
	because of the form of the log-partition function. Defined as a sum over $2^p$ terms, this function can not be computed in a reasonable time for $p\geq20$ so, for instance, maximum likelihood estimation can not be performed. 		
	
	One popular strategy to get around this problem has been proposed by \citet*{wainwright2007} and will be referred to as SepLogit in the sequel. From (\ref{bingraph_logist}), it directly follows that 
	parameters of model (\ref{Ising}) can be estimated through $p$ logistic regression models. More precisely, denote by $L(\btheta_j; \bcU_j, \bcU_{-j}) $ the log-likelihood under the logistic regression model (\ref{bingraph_logist}). Here $\bcU_j=(U_{j,1}, \ldots, U_{j,n}) \in \R^n$ contains the $n$ observations of variable $U_j$ and $ \bcU_{-j} \in \R^{n\times (p-1)}$ is the matrix containing the observations of the $p-1$ remaining variables, while $\btheta_j^T=(\theta_j, \btheta^T_{ -j})=(\theta_j, \theta_{j,1},..,\theta_{j,j-1},\theta_{j,j+1},...,\theta_{j,p})\in\R^p$ is the vector of parameters, over which optimization is performed to return (penalized) maximum likelihood estimates. Under SepLogit, estimation of $\btheta_j^*=(\theta^*_j,\btheta_{-j}^{*T})^T$ and selection of the non-zero values in $\btheta_{-j}^*$ are both achieved by minimizing the following lasso criterion, for an appropriate value of the regularization parameter $\lambda_j \geq 0$,
	\begin{equation}
		- L(\btheta_j; \bcU_j, \bcU_{-j}) + \lambda_j \|\btheta_{-j}\|_1.\label{SepLogit_Lasso}
	\end{equation}
	Here, $\|\btheta_{-j}\|_1 = \sum_{\ell\neq j} |\theta_{j,\ell}|$ is the $L_1$-norm of $\btheta_{-j}$.
	For every $j=1, ..., p$, we denote by $\hat{\btheta}_j$ any solution minimizing criterion (\ref{SepLogit_Lasso}). From these $p$ vectors  two estimates of the parameter $\theta^*_{j,\ell}$ are obtained for every $({j,\ell}) \in [p]^2$: $\hat{\theta}_{j,\ell}$ from the vector $\hat{\btheta}_j$ and  $\hat{\theta}_{\ell,j}$ from the vector $\hat{\btheta}_\ell$, with $\hat{\theta}_{j,\ell} \ne \hat{\theta}_{\ell,j}$ in general. Of course, it is even possible that $\hat{\theta}_{j,\ell}=0$ and $\hat{\theta}_{\ell,j} \ne 0$ for example. This asymmetry issue is resolved by considering either the SepLogit AND or the SepLogit OR strategy. According to the SepLogit AND strategy the edge $(j,\ell)$ is present in the edge set $E$ if both $\hat{\theta}_{j,\ell}$ and $\hat{\theta}_{\ell,j}$ are non-zero. According to SepLogit OR the edge $(j,\ell)$ is present in the set $E$ if either  $\hat{\theta}_{j,\ell}$ or $\hat{\theta}_{\ell,j}$ is non-zero.
	
	\subsection{Estimation of binary graphical models on \texorpdfstring{$K$}{} strata: existing approaches}
	In our illustrating example, clinicians expect injury tables to depend on the road user type (car occupants, pedestrians, ...). Therefore, associations among injuries suffered by victims of road accidents should be studied according to road user type. This means that we have to perform the estimation of $K$ binary graphical models, where $K$ is the number of road user types. In this context, we assume the existence of $K$ vectors  $\bTheta^{(k)*}=((\theta^{(k)*}_{j})_{1\leq j\leq p}, (\theta^{(k)*}_{j,\ell})_{1\leq j<\ell\leq p})^T$ in $\R^{p(p+1)/2}$, for $k=1, \ldots, K$, such that the probability of observing the injury table $\{\bU = \bu\}$ in the $k$-th stratum is given by
	\begin{equation}\label{Ising_K}
		\p_{\bTheta^{(k)*}}(\bU = \bu) = \exp\Big\{\sum_{j=1}^p\theta_j^{(k)*} u_j+
		\sum_{j=1}^{p-1}\sum_{\ell=j+1}^p\theta_{j,\ell}^{(k)*}u_ju_\ell
		- A(\bTheta^{(k)*})\Big\}.
	\end{equation}
	
	Of course, vectors $(\bTheta^{(k)*})_{k\in[K]}$ can be estimated separately, by applying the SepLogit method on each stratum independently. More precisely, set $\btheta_j^{(k)*}=(\theta_j^{(k)*},\btheta_{-j}^{(k)*T})^T=(\theta_j^{(k)*}, \theta_{j,1}^{(k)*},\\..,\theta_{j,j-1}^{(k)*},\theta_{j,j+1}^{(k)*},...,$ $\theta_{j,p}^{(k)*})^T\in\R^p$.  Estimates of these vectors returned by what we will refer to as Indep-SepLogit are defined as minimizers of the following criterion, for appropriate tuning parameters $\lambda_k \geq 0$,
	\begin{equation}
		\sum_{k=1}^{K}\left(-L(\btheta^{(k)}_{j}; \bcU^{(k)}_{j},\bcU^{(k)}_{-j}) + \lambda_k \|\btheta^{(k)}_{-j}\|_1\right), \label{SepLogit_Lasso_K}
	\end{equation}
	where $\bcU_j^{(k)}\in\R^{n_k}$ and $\bcU_{-j}^{(k)}\in\R^{n_k*(p-1)}$ contain the observations of variable $j$ and the $p-1$ remaining variables respectively, for the victims belonging to stratum $k$. Here $n_k$ is the number of observations belonging to the $k$-th stratum. 
	
	However, the main defect of Indep-SepLogit is that it does not account for the potential homogeneity among the vectors $\bTheta^{(k)*}$, $k\in[K]$. Indeed, even if associations between injuries may vary according to road user type, we still expect that $\theta^{(k_1)*}_{j,\ell} = \theta^{(k_2)*}_{j,\ell}$ for some $(k_1, k_2)\in[K]^2$ and some $(j,\ell)\in[p]^2$. By not accounting for this expected homogeneity, Indep-SepLogit exhibits two very undesirable properties. First, the returned estimates are of unnecessarily high dimension and so  typically have poor performance. Second, when homogeneity is expected, the identification of heterogeneities is generally of particular interest. In our example for instance, clinicians are interested in identifying which associations of injuries are more likely for each road user type; automobile manufacturers may further be interested in the associations of injuries that are more likely for car occupants, etc. However, differences between estimates $\hat \theta^{(k_1)}_{j,\ell}$  and $\hat \theta^{(k_2)}_{j,\ell}$ returned by Indep-SepLogit can not be interpreted as true differences, since  $\hat \theta^{(k_1)}_{j,\ell}$  and $\hat \theta^{(k_2)}_{j,\ell}$ are different by construction, as soon as these two quantities are non-null.

	In order to account for the expected homogeneity,  estimations of the vectors $(\bTheta^{(k)*})_{k\in[K]}$ have to be coupled in some way. \citet{Guo} proposed an approach based on the following multiplicative decomposition $\theta_{j,\ell}^{(k)*}=\phi^*_{j,\ell}\gamma_{j,\ell}^{(k)*}$. 
	Here, for all $j<\ell, $ $\phi^*_{j,\ell}$ is common to all $K$ strata and controls the occurrence of common links shared across strata, while $\gamma_{j,\ell}^{(k)*}$ is an individual factor specific to the $k$-th stratum, $k\in[K]$. The approach proposed by \citet{Guo} relies on the use of the pseudo-likelihood, which is another solution to get around the asymmetry issue of SepLogit. Moreover, the domain of parameters $\phi_{j,\ell}$ is restricted to  $\R_+$ to avoid sign ambiguities between $\phi_{j,\ell}$ and $\gamma_{j,\ell}^{(k)}$. More importantly, their approach relies on a penalty of the form $\eta_1 \sum_{j< \ell} \phi_{j,\ell} + \eta_2 \sum_{j<\ell}\sum_{k=1}^K |\gamma_{j,\ell}^{(k)}|$, as proposed by \citet*{zhou2010} under linear regression models. Keeping in mind that $\phi_{j, \ell}\geq 0$, the first term of the penalty shrinks estimates of  $\phi^*_{j,\ell}$ towards 0, while the second term shrinks estimates of $\gamma_{j,\ell}^{(k)*}$ toward 0. Therefore, the way this penalty couples the estimations across the $K$ strata is as follows: if $\hat \phi_{j,\ell} = 0$ then  $\hat \theta_{j,\ell}^{(k)} =0$ for all $k\in[K]$, and hence there is no link between nodes $j$ and $\ell$ in any of the $K$ graphs. On the other hand,  if $\hat \phi_{j,\ell} \neq 0$, then some of the $\hat\gamma_{j,\ell}^{(k)}$ and hence some of the $\hat \theta_{j,\ell}^{(k)}$ still have the possibility of being zero, for some $k\in[K]$. The coupling offered by this approach appears moderate. In particular, differences between non-zero estimates $\hat \theta^{(k_1)}_{j,\ell}$ and $\hat \theta^{(k_2)}_{j,\ell}$ still cannot be interpreted as true differences since $\hat \theta^{(k_1)}_{j,\ell}$ and $\hat \theta^{(k_2)}_{j,\ell}$ are different by construction, as long as they are both non-null. Therefore, this approach is not well suited for the application we have in mind.
	
	\subsection{Joint estimation of binary graphical models on \texorpdfstring{$K$}{} strata: our proposal}
	
	To fully account for the expected homogeneity across the $K$ graphs and then be able to interpret differences between estimates of, say, $\hat \theta^{(k_1)}_{j,\ell}$ and $\hat \theta^{(k_2)}_{j,\ell}$, our proposal relies on the combination of SepLogit and penalties used in the context of regression modeling on stratified data. In the first two following paragraphs, the principle of the estimation of the vectors $(\btheta^{(k)*}_j)_{k\in[K]}$, for a given $j\in[p]$ is presented, using either a fused penalty or the ideas of the data shared lasso \citep*{gross2016, ollier2017}. Next, we propose two strategies to combine the estimates obtained for all $j\in[p]$, following the ideas of SepLogit AND and SepLogit OR described above.  
	
	\subsubsection{Fused-SepLogit}
	
	Our first proposal follows the ideas developed in \citet*{danaher2014} under Gaussian graphical models as well as those of \citet*{tutz2010, tutz2012} and \citet{ViallonFused} under generalized linear regression models. It relies on a generalized fused lasso penalty. More precisely, for all $j\in [p]$, the method we will refer to as Fused-SepLogit returns estimates of $\btheta^{(k)*}_j$, for $k\in [p]$, defined as minimizers of the following criterion, for appropriate values of the tuning parameters $\lambda_1\geq 0$ and $\lambda_2 \geq 0$,
	\begin{equation}\label{FusedSepLogit}
		\sum_{k=1}^{K}\left(-L(\btheta^{(k)}_{j}; \bcU^{(k)}_{j},\bcU^{(k)}_{-j}) + \lambda_1 \|\btheta^{(k)}_{-j}\|_1\right) + \lambda_2\sum_{k_1<k_2}\|\btheta_j^{(k_1)}-\btheta_j^{(k_2)}\|_1.
	\end{equation}
	The fused-like penalty term $\|\btheta_j^{(k_1)}-\btheta_j^{(k_2)}\|_1$ shrinks estimates $\hat\btheta_j^{(k_1)}$ and $\hat\btheta_j^{(k_2)}$ towards each other, and therefore encourages equality of these two vectors. Accordingly, differences between components of $\hat\btheta_j^{(k_1)}$ and $\hat\btheta_j^{(k_2)}$ can be interpreted as true differences, and the expected homogeneity is likely to be fully accounted for.  More precisely, estimates derived from the adaptive version of (\ref{FusedSepLogit}) have been shown to enjoy an asymptotic oracle property, in the fixed $Kp$ case; \cite[see][]{tutz2012, ViallonFused}.

	\subsubsection{DataShared-SepLogit}
	Our second proposal consists in extending a method that was independently developed in \citet*{gross2016} and \citet*{ollier2017} under generalized linear models (see also \citet*{Viallon_AutoArxiv}). Applied to our context, it first relies, on the following additive decomposition of $\btheta_j^{(k)*}$
	\begin{equation}
		\btheta_j^{(k)^*}= \bmu^*_j + \bgamma_j^{(k)*}, \quad{\rm for \ each\ } k\in[K]
	\end{equation}
	where $\bmu^*_j\in\R^p$ ``morally'' contains what is common between the $K$ strata, while $\bgamma_j^{(k)^*}\in\R^p$ for $k\in [K]$ captures the variation in stratum $k$ around $\bmu^*_j$. Estimates of  $\bmu^*_j$ and the $\bgamma_j^{(k)*}$'s are then derived as minimizers of the following criterion, for appropriate values of $\lambda_1\geq 0$ and $\lambda_{2,k} \geq 0$,
	\begin{equation}
		\sum_{k=1}^{K}-L((\bmu_j + \bgamma_j^{(k)}); \bcU^{(k)}_{j},(\bcU^{(k)}_{-j})) + \lambda_1 \|\bmu_{j, -1}\|_1 + \sum_{k=1}^{K}\lambda_{2,k}\|\bgamma_j^{(k)}\|_1. \label{DataSharedSepLogit}
	\end{equation}
	It is easily shown that, at optimum, for all $\ell>1$, we have $\hat \mu_{j,\ell} = \argmin_{m\in\R} (\lambda_1 |m| + \sum_k \lambda_{2,k}|\hat\theta^{(k)}_{j, \ell}-m|$), and is therefore a weighted and shrunk towards 0 version of the median of $(\hat\theta^{(1)}_{j, \ell}, \ldots, \hat\theta^{(K)}_{j, \ell})$. For the constant terms, we have $\hat \mu_{j,1} = \argmin_{m\in\R} \sum_k \lambda_{2,k}|\hat\theta^{(k)}_{j}-m|$ and $\hat \mu_{j,1}$ is then simply a weighted median of  the set of values $(\hat\theta^{(1)}_{j},\ldots, \hat\theta^{(K)}_{j})$.  In other words, the penalty term 
	$\sum_{k=1}^{K}\lambda_{2,k}\|\bgamma_j^{(k)}\|_1=\sum_{k=1}^{K}\lambda_{2,k}\|\btheta_j^{(k)} - \bmu_j\|_1$ shrinks the estimators $\hat\btheta_j^{(k)}$, $k\in[K]$, towards their ``weighted and shrunk towards 0'' median. Sparsistency of the approach has been studied in a non-asymptotic framework in \citet*{ollier2017}. In particular, data shared lasso is shown to generally outperform the more standard strategy based on an a priori selection of a reference stratum, for roughly the same computational cost. Indeed, data shared lasso can be rewritten as a standard lasso (under logistic regression models here), after a simple transformation of the original data. We refer to \citet*{ollier2017} and \citet*{gross2016} for more details.

	\subsubsection{Combining the \texorpdfstring{$(\hat \btheta_j^{(k)})_{j\in[p], k\in[K]}$}{} to derive the \texorpdfstring{$K$}{} estimated graphs}

	For every $j<\ell$, both Fused-SepLogit and DataShared-SepLogit return two vectors of estimates for $(\theta_{j,\ell}^{(1)*}, ..., \theta_{j,\ell}^{(K)*})$: $(\hat{\theta}_{j,\ell}^{(1)}, ..., \hat{\theta}_{j,\ell}^{(K)})$ and  $(\hat{\theta}_{\ell,j}^{(1)}, ..., \hat{\theta}_{\ell,j}^{(K)})$. Of course, we may still have, for some $k, \ell, j$, $\hat {\theta}_{j,\ell}^{(k)}\neq \hat{\theta}_{\ell,j}^{(k)}$, or even $\hat{\theta}_{j,\ell}^{(k)}=0$ while $\hat{\theta}_{\ell,j}^{(k)}\neq 0$ for instance. But we may also have other asymmetry issues here. For instance, we may have  a fully homogeneous vector $(\hat{\theta}_{j,\ell}^{(1)}, ..., \hat{\theta}_{j,\ell}^{(K)})$, that is, a vector whose components are all equal suggesting that the association between variables $U_\ell$ and $U_j$ does not vary across the strata, while $(\hat{\theta}_{\ell,j}^{(1)}, ..., \hat{\theta}_{\ell,j}^{(K)})$ exhibits some heterogeneities, suggesting that the association between variables $U_\ell$ and $U_j$ does vary across the strata. 
	
	To get around these asymmetry issues, we propose two strategies, referred to as (Fused or DataShared)-SepLogit MIN and (Fused or DataShared)-SepLogit MAX, which can be seen as adaptations of SepLogit AND and SepLogit OR to our context. When only one graph has to be estimated on a single stratum (or on the population as a whole), the complexity of the estimated graph can be defined as the number of edges of this graph. Then SepLogit AND [resp. OR] returns the graph with lowest [resp. highest] complexity given the vectors $\hat \btheta_j$. When $K$ graphs are estimated on $K$ strata, our two strategies, MIN and MAX, also return two graphs, with lowest and highest complexities, respectively. But the definition of complexity has to be adapted. For every $(j<\ell)$, denote by $\boldsymbol{S}_j$ the vector $(\hat{\theta}_{j,\ell}^{(1)},...,\hat{\theta}_{j,\ell}^{(K)}) \in \R^K$ and by $\boldsymbol{S}_{\ell}$  the vector $(\hat{\theta}_{\ell,j}^{(1)},...,\hat{\theta}_{\ell,j}^{(K)}) \in \R^K$ returned by either Fused-SepLogit or DataShared-SepLogit. Then, for any vector $\boldsymbol{S}=(s_1,...,s_K)\in\R^K$, we define its complexity $comp(\boldsymbol{S})$ as the number of distinct non-null values among $(s_1,...,s_K)$.
	
	For every $(j<\ell)\in[p^2]$, let $\boldsymbol{S}_{j,\ell}^{\min}$ [resp. $\boldsymbol{S}_{j,\ell}^{\max}$] denote the vector among $(\boldsymbol{S}_j, \boldsymbol{S}_\ell)$ with lowest [resp. highest] complexity, as measured by function $comp$. In other words, if $comp(\boldsymbol{S}_j) < comp(\boldsymbol{S}_\ell)$ then $\boldsymbol{S}_{j,\ell}^{\min} = \boldsymbol{S}_j$ and $\boldsymbol{S}_{j,\ell}^{\min}=\boldsymbol{S}_\ell$. (Fused or DataShared)-SepLogit MIN [resp. MAX] then returns graphs constructed from vectors $\boldsymbol{S}_{j,\ell}^{\min}$ [resp. $\boldsymbol{S}_{j,\ell}^{\max}$] obtained for all $j<\ell$.

	\section{Simulation study}\label{sec:Simul}
	We empirically compared the approaches presented above on synthetic data,  following the simulation framework developed in \citet{Guo}. Indep-SepLogit and DataShared-SepLogit were implemented using the glmnet package, while Guo and others' approach and Fused-SepLogit were implemented using the BMNPseudo and FusedLasso packages respectively. Note that the FusedLasso package is not maintained on the CRAN anymore, but is still available from the archives. For the sake of brevity, results are only presented for the MIN strategy presented above. Selection of tuning parameters was performed using the BIC. Following ideas introduced in \citet{Lars} \cite[see also][]{ViallonFused, Viallon_AutoArxiv}, a two-step BIC approach was also considered and yielded very similar performance (results not shown).

	\subsection{Data generation}
	
	The IsingSampler package of R was used to generate the data, given matrices $\bTheta^{(k)*}=(\theta_{j,\ell}^{(k)*})_{p*p}$. Following the framework considered by \citet{Guo}, these matrices were defined as $\bTheta^{(k)*}=\bmu^*+\boldsymbol{\Psi}^{(k)*}$, where $\bmu^*=(\bar{\mu}^*_{j,\ell})_{p*p}$ represents the common structure across all strata and $\boldsymbol{\Psi}^{(k)*}=(\psi_{j,\ell}^{(k)*})_{p*p}$ represents the structure specific to stratum $k$, for $k\in [K]$. 
	
\begin{figure}[!ht]
			\centering
			
			\includegraphics{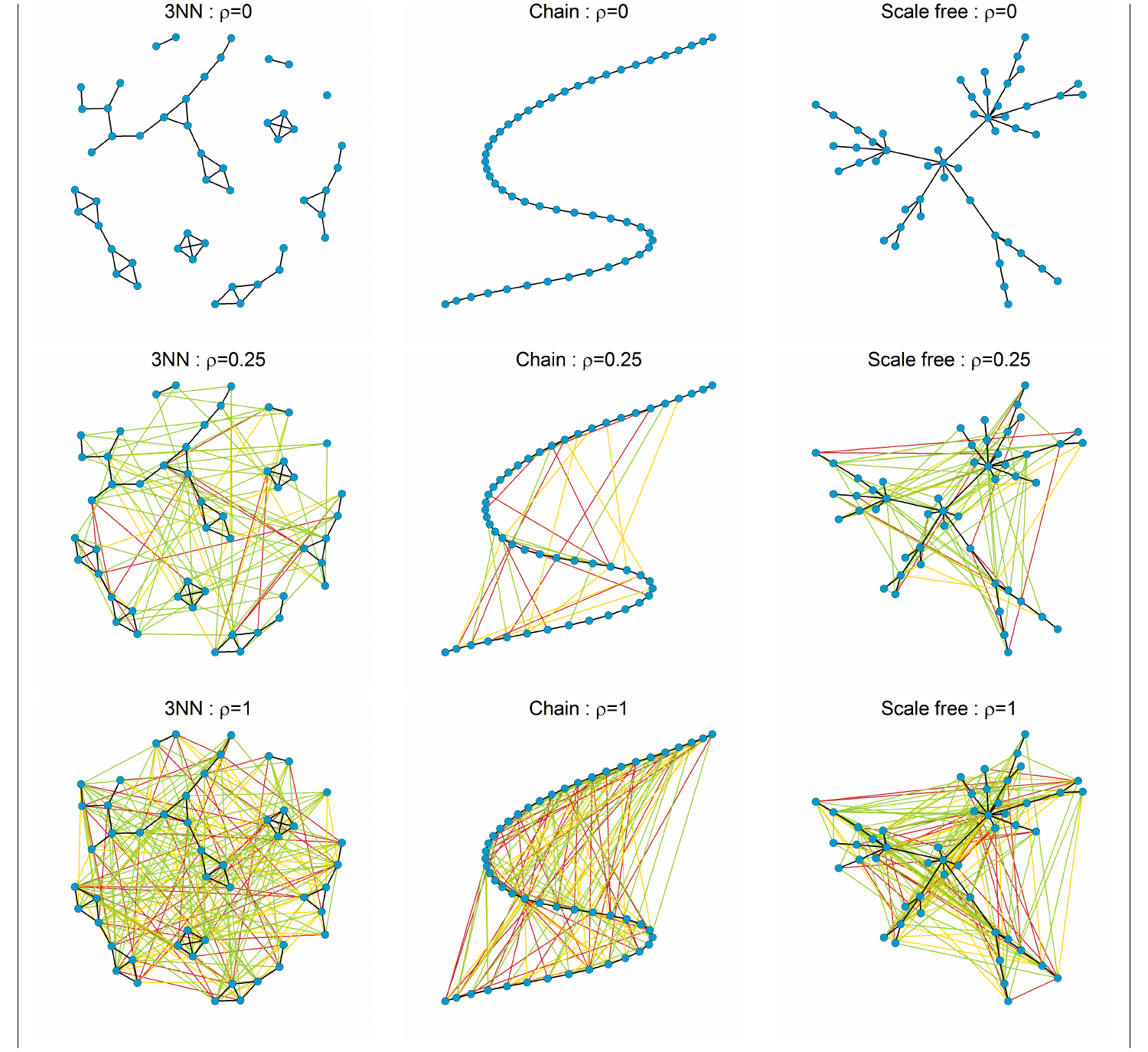}
			\caption{A graphical representation for the three types of network of structuring with a ratio $\rho$ equal to 0, 0.25 and 1, with $p=50$ and $K=3$. The black edges represent the common structure and the red, blue and green edges represent the structures specific to each stratum.}\label{CommonStructure_Ratio}
		\end{figure}

	For the common part, we again followed the framework of \citet{Guo} and considered three types of graphs which are the chain graph, the 3-nearest neighbor graph and the scale-free graph. See Figure \ref{CommonStructure_Ratio} and \citet{Guo} for more details. \vspace{0.1cm}
	%\paragraph{Trois types de graphes} on a utilisé les trois types de graphes suivants:\\
	As for the specific part, non-zero values of  each $\boldsymbol{\Psi}^{(k)*}$ are randomly generated on the set $[$$-1$,$-0.5$$]\cup[0.5,1]$. The number of non-zero values in each $\boldsymbol{\Psi}^{(k)*}$ is the same for every $k$ and depends on the parameter $\rho\in[0, 1]$. This parameter corresponds to the ratio between the number of individual edges and the number of common edges. Therefore, this ratio represents the level of heterogeneity between the different strata. In particular, if $\rho=0$, the structures are identical for the $K$ strata. In this study, five values of $\rho$ were considered: 0, 0.25, 0.5, 0.75 and 1.

	For each common structure and ratio $\rho$, we considered 50 replicates of data consisting in 500 observations in each stratum, with $(p=10, K=3)$ in a first simulation study and $(p=40, K=4)$ in the second one. These choices were motivated by the dimension in our leading example (see Section \ref{sec:Registre}) and also by the slowness of both Fused-SepLogit and the approach of \citet{Guo}, which were only run on the first simulation study to save computational time.

	\subsection{Evaluation criteria}
	Estimates $\hat{\bTheta}^{(k)}=(\hat{\theta}_{j,\ell}^{(k)})_{p*p}$  of each $\bTheta^{(k)*}$ returned by each method were computed and compared to $\bTheta^{(k)*}$ on each simulated data. Two types of criteria were computed and averaged over the 50 replicates of each of the considered simulation design. The first one, Acc.S, measures the accuracy regarding the support of each matrix $\bTheta^{(k)*}$, that is the identification of the non-zero values among the off-diagonal elements of matrices $\bTheta^{(k)*}$, $k\in[K]$. More precisely, it is defined as
	$$\text{Acc.S}=\frac{1}{K}\sum_{k\in[K]}\left(\frac{\sum_{j>\ell}\left(\mathds{1}[\theta_{j,\ell}^{(k)*}\neq0\hspace{0.1cm},\hspace{0.1cm}\hat{\theta}_{j,\ell}^{(k)}\neq0]+\mathds{1}[\theta_{j,\ell}^{(k)*}=0\hspace{0.1cm},\hspace{0.1cm}\hat{\theta}_{j,\ell}^{(k)}=0]\right)}{p(p-1)/2}\right).$$
	
	We also evaluated the performance of each method regarding the identification of the heterogeneity between matrices $\bTheta^{(k)*}$, $k\in[K]$. Here, we report results for Acc.H, which is defined as follows
	$$
	\text{Acc.H}=\frac{\sum_{j>\ell}\left(\mathds{1}[Z^*_{j,\ell}\neq0\hspace{0.1cm},\hspace{0.1cm}\hat{Z}_{j,\ell}\neq0]+\mathds{1}[Z^*_{j,\ell}=0\hspace{0.1cm},\hspace{0.1cm}\hat{Z}_{j,\ell}=0]\right)}{p(p-1)/2}, 
	$$
	where 
	
	$$ Z^*_{j,\ell} =
	\begin{cases}
	0       & \quad \text{if\space\space} \theta_{j,\ell}^{(1)^*}=\theta_{j,\ell}^{(2)^*}=...=\theta_{j,\ell}^{(K)^*}\\
	1  & \quad \text{otherwise}
	\end{cases}
	\hfill {\rm and} \quad
	\hat{Z}_{j,\ell} =
	\begin{cases}
	0       & \quad \text{if\space\space} \hat{\theta}_{j,\ell}^{(1)}=\hat{\theta}_{j,\ell}^{(2)}=...=\hat{\theta}_{j,\ell}^{(K)}\\
	1  & \quad \text{otherwise}.\\
	\end{cases}
	$$
	In other words, $Z^*_{j,\ell}$ [resp. $\hat Z_{j,\ell}$] is a binary variable which equals 1 if the association between variables $j$ and $\ell$ vary across the $K$ strata under the considered simulation design, under the true model [resp. as identified by the considered method], and Acc.H corresponds to the accuracy regarding the support recovery of $Z^*_{j,\ell}$. Other criteria, such that the Rand index, were considered and lead to very similar results (not shown). 
	
	\subsection{Results}
	\subsubsection{First simulation study with \texorpdfstring{$p=10$}{}, \texorpdfstring{$K=3$}{} and \texorpdfstring{$n_k=500$}{}}
		\begin{figure}[!ht]
			\centering
			\includegraphics[width=1\linewidth]{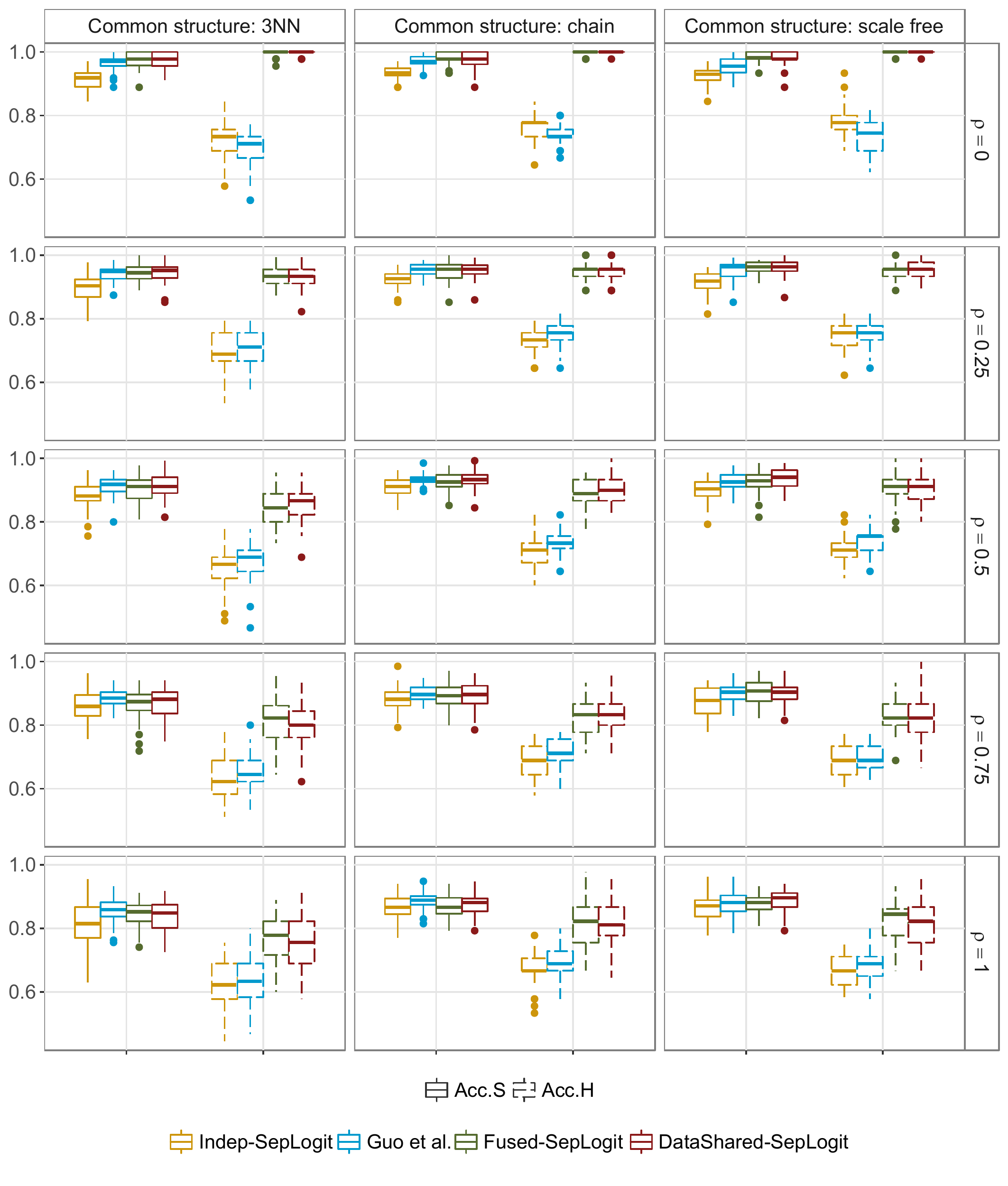}
			\caption{Boxplots for the values of Acc.S and Acc.H obtained for each method on the 50 replicates of each simulation design in the first simulation study.}
			\label{Results_SIM1}
		\end{figure}
	Results are presented on Figure \ref{Results_SIM1}. A first remark is that the type of the common structure  only marginally affects the comparison between the four methods. First consider Acc.S. Overall, the performance of Indep-SepLogit is independent of the level of heterogeneity $\rho$, while the other three methods globally outperform Indep-SepLogit and are all the better as $\rho$ is low, as expected since they all account for homogeneity when it is present. These three methods share similar performance, with a slight advantage for Fused-SepLogit and DataShared-SepLogit if $\rho=0$ and an even slighter advantage for Guo and others' method for $\rho\geq 0.75$. 
	As for Acc.H, Fused-SepLogit and DataShared-SepLogit perform similarly and noticeably outperform both Indep-SepLogit and Guo and others' method, especially for low values of $\rho$. In addition, Guo and others' method and Indep-SepLogit perform similarly regarding this criterion confirming that Guo and others' method is not well suited for the identification of heterogeneities.

	\subsubsection{Second simulation study with \texorpdfstring{$p=40$}{}, \texorpdfstring{$K=4$}{} and \texorpdfstring{$n_k=500$}{}}
	
	\begin{figure}[!ht]
		\centering
		\includegraphics[width=1\linewidth]{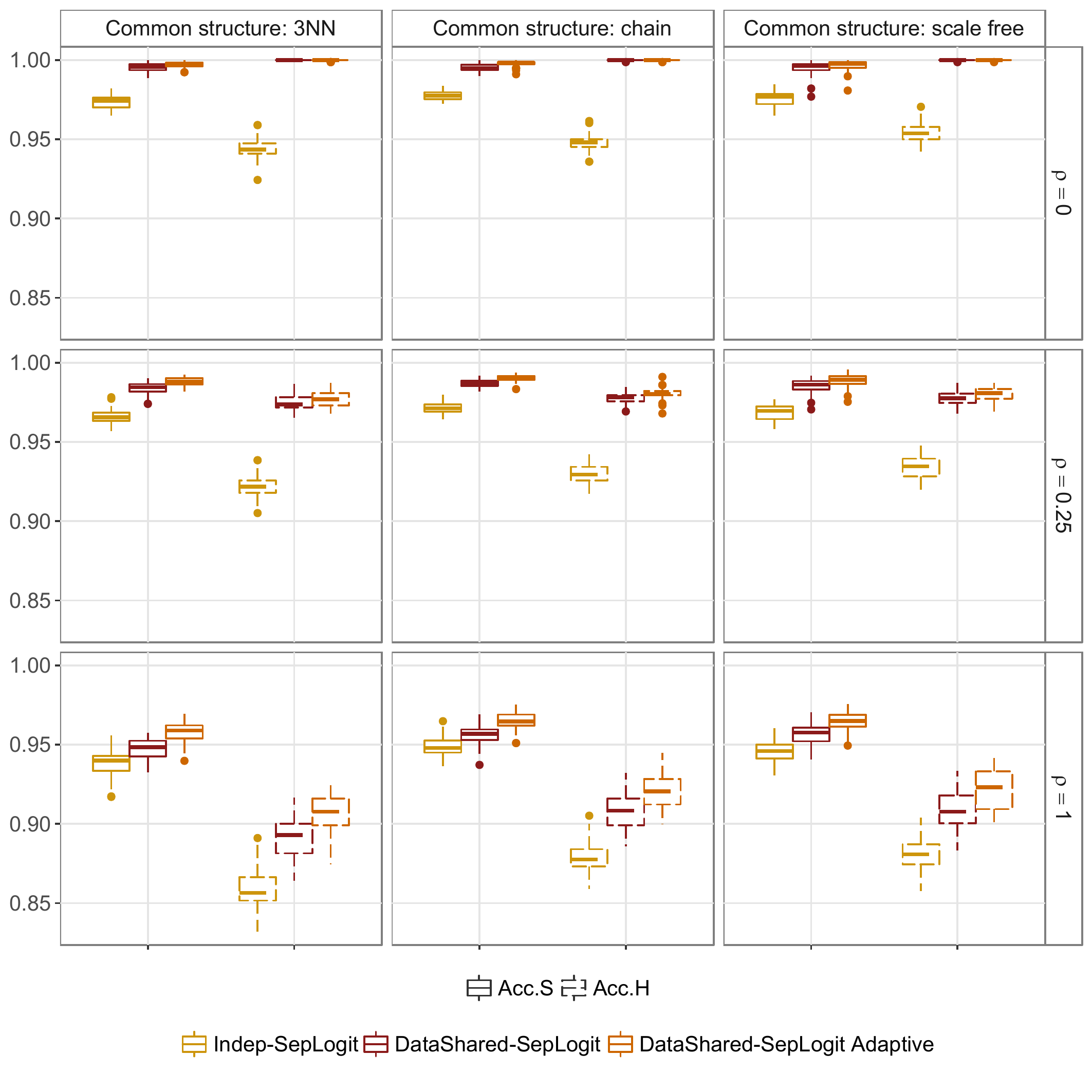}
		\caption{Boxplots for the values of Acc.S and Acc.H obtained for each method on the 50 replicates of each simulation design in the second simulation study.}
		\label{Results_SIM2}
	\end{figure}
	
	Motivated by the dimensions involved in our illustrative example, we further considered the situation where $K=4$ and $p=40$. However, both Guo and others' method and Fused-SepLogit were too slow and only results for Indep-SepLogit and DataShared-SepLogit are presented. We also considered the adaptive version of DataShared-SepLogit, using unpenalized maximum likelihood estimates to construct the adaptive weights; (see \cite{Viallon_AutoArxiv} for details on the adaptive version of data shared lasso). Results are presented on Figure \ref{Results_SIM2}. They are consistent with those obtained in the previous simulation study. More precisely, DataShared-SepLogit outperforms Indep-SepLogit regarding the identification of the heterogeneities (as measured by Acc.H), as well as support recovery, especially for low values of $\rho$ (that is, for high level of homogeneity). The adaptive version of DataShared-SepLogit globally outperfoms the standard version regarding both Acc.S and Acc.H, in particular for high values of $\rho$.
	
	\section{Application}\label{sec:Registre}
\begin{figure}[!ht]
			\centering
			\includegraphics{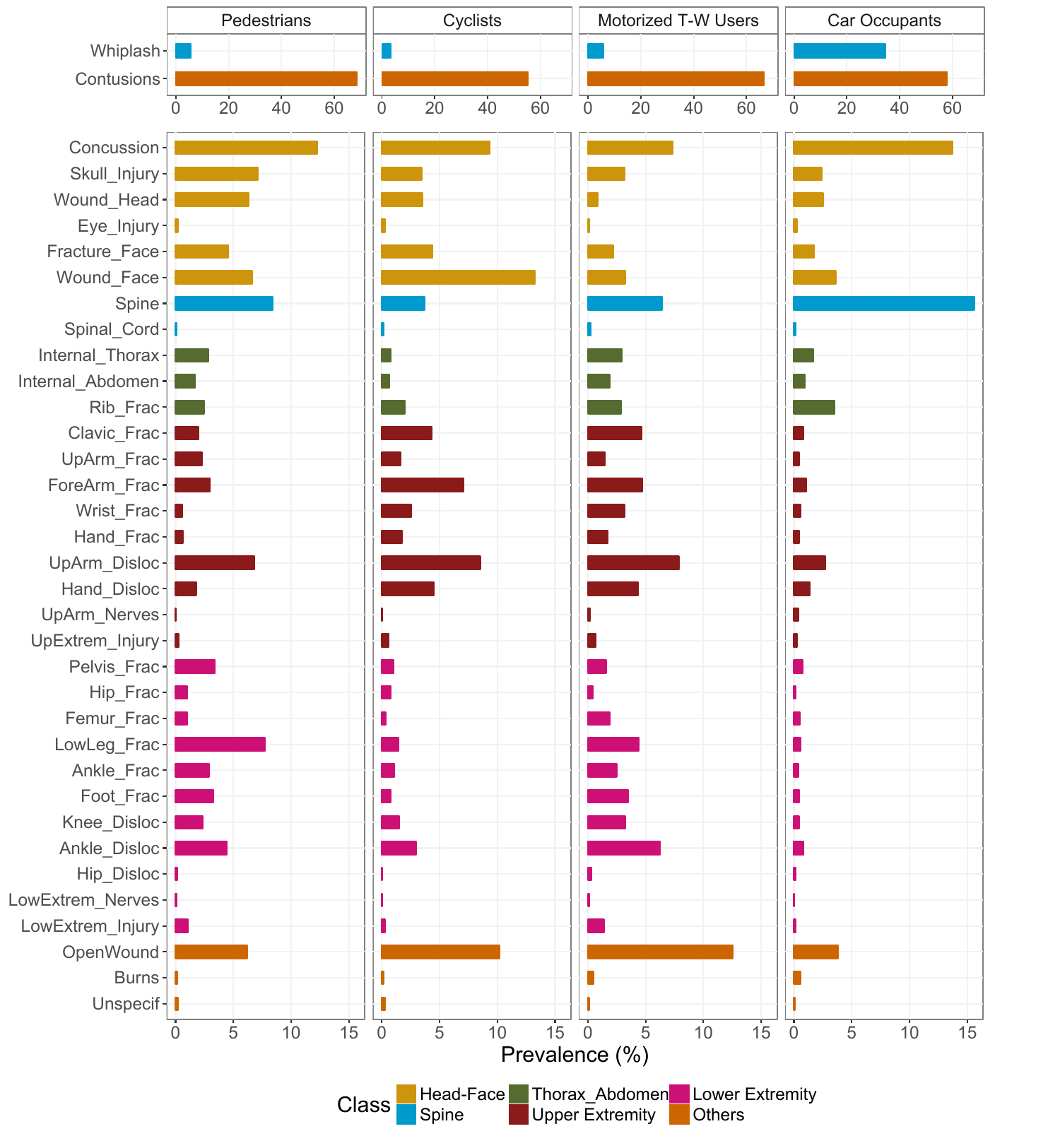}
			\caption{Injury prevalences in each stratum.}
			\label{InjuryPrevalences}
			
		\end{figure}

	\paragraph{Data description (Registry data)}
	The Rh\^one registry contains all the data of road traffic accidents victims in the Rh\^one. Rh\^one is a French Department (1,600,000 inhabitants) in the Rh\^one-Alpes Region whose main urban center is Lyon. This registry is officially recognized by the National Registry Committee and it is managed by the ARVAC (Association pour le Registre des Victimes d'ACcidents de la route). 
	Data collection began on January, 1st 1995. Up to now, the data from 1996 to 2014 have been fully computerized and validated. These data contain the identification of the victims (name, sex, date of birth), administrative information (address, accident at work), the accident characteristics (date, time, place, type of vehicle, etc.) as well as the complete injury tables of victims. Injury tables contains each injury suffered by the victim, coded according to the \textbf{A}bbreviated \textbf{I}njury \textbf{S}cale \textbf{90} classification. This classification describes the type and the location of the injury using six digits written as 12(34)(56). The first digit identifies the body region \textbf{[R]} (Head, Face, Neck, Spine, etc.), the second one identifies  the type of anatomic structure \textbf{[T]} (Whole Area, Vessels, Nerves, etc.),  the third and fourth digits identify the specific anatomic structure, or the nature of the injury when an entire area is reached \textbf{[S]} (Cervical Column, Back Column, Contusion, Burn, etc.) and the fifth and sixth ones specify the type of injury \textbf{[N]} (Fractures, rupture, laceration, etc.). Using this coding, the original dataset contained 1348 distinct codes corresponding to 1348 distinct injuries. However, because most of these injuries have very low prevalences, we decided to group some of them. After converting the AIS codes to ICD-10 codes (International Statistical Classification of Diseases and Related Health Problems, $10$-th Revision), we used the grouping proposed in the EUROCOST model \cite[see][Table 1]{lyons2006}. We further decided to split the "Internal-organ injury" group into two groups, corresponding to "Internal-organ injury thorax" and "Internal-organ injury abdomen". We finally worked with 36 groups of injuries. Formally, this led us to consider 36 binary $(0,1)$-variables $U_1, \ldots, U_p$, where $U_j=1$ if and only if the original injury table of the victim contains at least one injury falling into the $j$-th group of the EUROCOST model.  Further note that these 36 groups of injuries can be classified into 6 classes, roughly corresponding to body areas (see Table \ref{Injury Group}). From now on the 36 groups of injuries will simply be referred to as injuries. \\	
	In this application, we will use the data of the last 10 years (2005 to 2014) on 4 strata: pedestrians, cyclists, motorized two-wheelers users and car occupants. Overall, these data contain 67,894 victims and 109,793 injuries. See Table \ref{Data} for more details. In Figure \ref{InjuryPrevalences}, we present the prevalence of each injury on the different strata. Colors correspond to the class of each injury and the length of each bar corresponds to the prevalence of each injury.  Some differences can be noticed across the strata. For instance, compared to other road users, car occupants are more likely to suffer from whiplash and spine injuries, while they are less likely to suffer from injuries in either upper or lower extremities. Motorized two wheelers users are less likely to suffer concussion and wound face, as expected since they are supposed to wear a helmet.

	\begin{table}
		\caption{Descriptions, labels and classes of injuries
			\label{Injury Group}}
		\centering
	\resizebox{\linewidth}{!}{	
		\begin{tabular}{lll}
			\toprule
			Description   &	Label &	Class	\\ \midrule
			Concussion &	Concussion &	Head-Face\\
			Other skull-brain injury &	Skull-Injury &	Head-Face  	\\
			Open wound of head&	Wound-Head&	Head-Face\\
			Eye injury&	Eye-Injury&	Head-Face\\
			Fracture of facial bones&	Fracture-Face&	Head-Face\\
			Open wound of face&	Wound-Face&	Head-Face\\[3pt]
			Fracture/dislocations/sprain/strain of vertebral/spine&	Spine&	Spine\\
			Whiplash injury/neck sprain/distortion of cervical spine 	&	Whiplash&	Spine\\
			Spinal cord injury&	Spinal-Cord&Spine\\[3pt]
			Internal-organ injuries /Thorax&	Internal-Thorax	&	Thorax-Abdomen	\\
			Internal-organ injuries /Abdomen&	Internal-Abdomen	&	Thorax-Abdomen	\\
			Fracture of rib/sternum&	Rib-Frac	&	Thorax-Abdomen	\\[3pt]
			Fracture of clavicle or scapula&	Clavic-Frac	&	Upper Extremity\\
			Fracture of upper arm&	UpArm-Frac	&	Upper Extremity\\
			Fracture of elbow or forearm&	ForeArm-Frac	&	Upper Extremity\\
			Fracture of wrist&	Wrist-Frac		&Upper Extremity\\
			Fracture of hand or fingers&	Hand-Frac	&	Upper Extremity\\
			Dislocation/sprain/strain of shoulder or elbow	& UpArm-Disloc	 &	Upper Extremity\\
			Dislocation/sprain/strain of hand or fingers&	Hand-Disloc		& Upper Extremity\\
			Injury to nerves of upper extremity	& UpArm-Nerves	 &	Upper Extremity\\
			Complex soft tissue injury of upper extremity&	UpExtrem-Injury	& Upper Extremity\\[3pt]
			Fracture of pelvis	& Pelvis-Frac	 &	Lower Extremity\\
			Fracture of hip	& Hip-Frac	&	Lower Extremity\\
			Fracture of femoral shaft &	Femur-Frac	& Lower Extremity\\
			Fracture of knee or lower leg&	LowLeg-Frac	 &	Lower Extremity\\
			Fracture of ankle&	Ankle-Frac	& Lower Extremity\\
			Fracture of foot(excludes ankle)	& Foot-Frac	&	Lower Extremity\\
			Dislocation/sprain/strain of knee&	 Knee-Disloc	&	Lower Extremity\\
			Dislocation/sprain/strain of ankle or foot	& Ankle-Disloc	 &	Lower Extremity\\
			Dislocation/sprain/strain of hip&	Hip-Disloc	&	Lower Extremity\\
			Injury to nerves of lower extremity	& LowExtrem-Nerves	 &	Lower Extremity\\
			Complex soft tissue injury of lower extremity&	LowExtrem-Injury	 &	Lower Extremity\\[3pt]
			Superficial injury (including contusions and bruises)&	Contusions	&	Others\\
			Open wounds	& OpenWound	 &	Others\\
			Mild burns	& Burns	&Others\\
			Other and unspecified injury&	Unspecif	&	Others\\
			\bottomrule
		\end{tabular}
	}
	\end{table}
	
		\begin{table}
			
			\caption{Number of victims and injuries in each stratum
				\label{Data}}
			\centering
			
			\begin{tabular}{lcccc}
				\toprule
				&	Pedestrians & Cyclists & Motorized T-W Users & Car Occupants	\\ \midrule
				
				Victims & 6663 &  11431 & 19204 & 30596 \\
				Injuries: before grouping  &	15348  &	20703 &	41779 & 55966\\
				Injuries: after grouping &	11865  &	17404 &	32932 & 47592\\
				\bottomrule
			\end{tabular}
			
		\end{table}

\begin{figure}[!ht]
			\centering
			\includegraphics[width=0.95\linewidth]{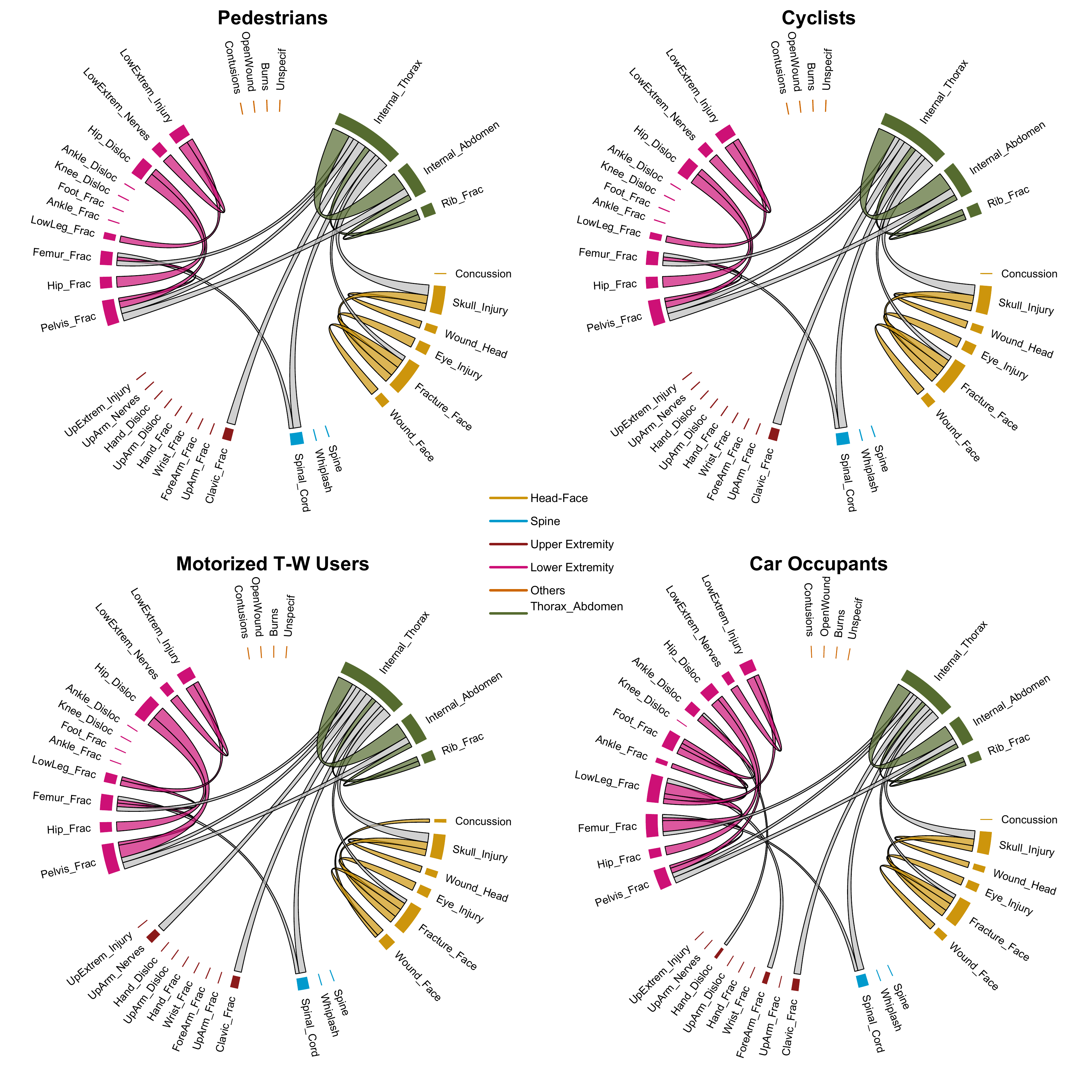}
			\caption{Application of the DataShared-SepLogit approach on the Rh\^one Registry Data to describe the injury tables of road accidents victims, according to road user type: pedestrians, cyclists, motorized T-W and car occupants. Only edges corresponding to conditional odds-ratios greater than or equal to 2 are represented.}
			\label{App_DataSharedSepLogit_log(2)}
		\end{figure}

	\paragraph{Stratified graphical model estimation}
	When applied to this data set, Fused-SepLogit and Data Shared-SepLogit led to very similar structures, as was expected from the results of the simulation study described in Section \ref{sec:Simul}.  For the sake of brevity, only results obtained with DataShared-SepLogit MIN are presented here, where the BIC was used to select the regularization parameters.\\	
	Figure \ref{App_DataSharedSepLogit_log(2)} presents a simplified version of the estimated structure of the graphical models, on each of the four strata. For the sake of legibility, only edges corresponding to estimated conditional odds-ratios greater than or equal to two are represented (see Table \ref{NA} for more details, and the Discussion for our motivation to focus on these associations only).  The chordDiagram function of the circlize R package was used to generate this figure. Nodes (injuries) are represented on the circle to facilitate the comparison of the structures across the strata. Colors of the nodes correspond to the class they belong to (see Table \ref{Injury Group}). The color of each edge is related to the classes of the two injuries it connects. If these two injuries belong to distinct classes, then the edge is gray, otherwise the edge shares the same color as the two injuries. The width of each edge is related to the value of the corresponding conditional odds-ratio. Lastly, the size of each node is the sum of the edge widths over the edges involving this node. \\
	Overall, the estimated structures are very similar over the four strata, and most edges connect injuries belonging to the same class. In particular, associations between injuries of the Head-Face class are very similar for motorized two-wheelers users and the other users. This can be considered as unexpected, given that the helmet is supposed to protect these users against injury in this area: indeed, Figure \ref{InjuryPrevalences} shows that injuries are less frequent  in the Head-Face area for motorized two-wheelers users. Our findings actually suggest that if a motorized two-wheelers user suffers from one injury in the Head-Face area, then it usually implies that the helmet was not able to protect this user well enough (either because of the violence of the impact, or because the helmet felt down, or because the user simply did not use any helmet, etc.), and that this user is likely to suffer from other injuries in the Head-Face area as well, just as any other road user. \\	
	On the other hand, the main difference across the four estimated structures concerns associations between injuries in the Lower Extremity area, which are more numerous for car occupants, while prevalences of these injuries are lower for these road users (see Figure \ref{InjuryPrevalences}). A possible explanation is that such injuries  for car occupants are likely to be due to substantial deformations of the car body or collisions with the dashboard, which are likely to generate multiple injuries.  
		\begin{table}[!ht]
			\caption{Number of associations in each stratum
				\label{NA}}
			\centering
			\begin{tabular}{lcccc}
				\toprule
				Odds-ratio $\backslash $ Stratum &	Pedestrians & Cyclists & Motorized T-W Users & Car Occupants	\\ \midrule
				
				$\neq 1$ & 237 &  264 & 242 & 247 \\
				$>1$  &	39  &	37 &	43 & 45\\
				$\geq 2$  &	20  & 20 &	23 & 26\\
				\bottomrule
			\end{tabular}
		\end{table}

\section{Discussion}\label{sec:Disc}
In this article, we described two methods based on multiple penalized logistic regressions to jointly estimate binary graphical models on several pre-defined strata. By appropriately penalizing heterogeneities across the corresponding structures, the proposed methods take benefit of the potential homogeneity among these structures, and allow the interpretation of the identified heterogeneities. Focusing on the identification of heterogeneities, we observed better performance for our methods 
compared to both the naive strategy consisting in estimating the structure on each stratum independently, and the strategy proposed by \citet{Guo}. We may further mention that methods based on group lasso penalties, such as that of  \citet*{ma2016}, were not considered in this work since they are not well suited for the identification and interpretation of heterogeneities either. Other potential strategies that allow the interpretation of the identified heterogeneities include the method developed in \citet{cheng2014} to account for covariates in the Ising model. In our context, these covariates would be indicators of the strata, after the selection of one reference stratum $r \in[K]$. This strategy would rely on the reparametrization $\theta_{j,\ell}^{(k)}=\theta_{j,\ell}^{(r)}+\gamma_{j,\ell}^{(k)}$, if $k\neq r$, and would penalize both the $|\theta_{j,\ell}^{(r)}|$'s and the $|\gamma_{j,\ell}^{(k)}|$'s. As mentioned in \citet*{ollier2017} and \citet*{gross2016}, this strategy is close in spirit to data shared lasso, the latter having the clear advantage of bypassing the arbitrary choice of the reference stratum and of mimicking the performance one would achieve by selecting an optimal reference stratum (which may vary for each association $j, \ell$). For instance, in our application, epidemiologists may be tempted to apply this strategy with car occupants as the reference stratum, which appears as a ``natural'' reference stratum since it is the largest stratum. Figure \ref{App_RefLassoSepLogit_log(2)} in the Appendix presents the results obtained by doing so, focusing on associations corresponding to conditional odds-ratios greater than or equal to 2 again. This approach identifies fewer heterogeneities than ours, as expected from the results of \citet*{ollier2017} since most of the heterogeneities that were identified by DataShared-SepLogit concerned car occupants; see the Appendix for more details. Lastly, we may mention the work of \citet{tao2016} which uses $\ell_0$-like penalties to encourage homogeneity when estimating multiple Gaussian graphical models. Extending their approach to binary graphical models would be an interesting lead for future work. \\
In the application of our methods to describe the injury tables of victims of road accident according road user types, we observed that most associations were common to all user types, while heterogeneities mostly concerned car occupants. We have to insist on the fact that only positive associations were presented here (Figure \ref{App_DataSharedSepLogit_log(2)} actually presents associations corresponding to conditional odds-ratio greater than or equal to two only; see Figure \ref{App_DataSharedSepLogit_log(1)} in the Appendix for models with the whole list of positive associations). The main reason why negative associations are less interesting in this particular application has to do with the way clinicians record injuries for each victim: most often, they neglect to record some injuries when more severe injuries are present, and most negative associations are simply an illustration of this recording bias. In addition, because all observations correspond to victims of road accidents, our data set consists of individuals who suffer from at least one injury. Therefore, our sample is biased compared to the whole population; the resulting  collider bias \citep*{Hernan} typically makes causal interpretation of negative association hazardous. We may however notice that describing associations in the injury tables of victims of road accidents is still relevant, since the sub-population of victims is the one clinicians have to take care of, even if no causal interpretation can be given to the identified associations. In future work, finer groupings of the injuries may be used to improve clinical interpretability, along with other definitions for the strata, which may include the severity of the crash, etc.  
Lastly, in applications where selection bias is absent, combining the ideas described here with those presented in \citet*{champion2017} (see also \citet*{vandeGeer} for the $\ell_0$ version) could lead to a powerful approach to estimate causal DAGs on stratified data.

\bibliographystyle{biorefs}
\bibliography{refs}

\begin{thebibliography}{99}

\bibitem[Banerjee \emph{and others}(2008)Banerjee, El~Ghaoui and
  d'Aspremont]{banerjee2008}
\textsc{Banerjee, O., El~Ghaoui, L. and d'Aspremont, A.} (2008).
\newblock Model selection through sparse maximum likelihood estimation for
  multivariate gaussian or binary data.
\newblock {\em Journal of Machine learning research\/}~\textbf{9}(Mar),
  485--516.

\bibitem[Besag(1974)Besag]{besag1974}
\textsc{Besag, J.} (1974).
\newblock Spatial interaction and the statistical analysis of lattice systems.
\newblock {\em Journal of the Royal Statistical Society. Series B
  (Methodological)\/}, 192--236.

\bibitem[Champion \emph{and others}(2017)Champion, Picheny and
  Vignes]{champion2017}
\textsc{Champion, M., Picheny, V. and Vignes, M.} (2017).
\newblock Inferring large graphs using $\ell_1$-penalized likelihood.
\newblock {\em Statistics and Computing\/}, 1--17.

\bibitem[Cheng \emph{and others}(2014)Cheng, Levina, Wang and Zhu]{cheng2014}
\textsc{Cheng, J., Levina, E., Wang, P. and Zhu, J.} (2014).
\newblock A sparse ising model with covariates.
\newblock {\em Biometrics\/}~\textbf{70}(4), 943--953.

\bibitem[Cox and Wermuth(1994)Cox and Wermuth]{cox1994}
\textsc{Cox, D.~R. and Wermuth, N.} (1994).
\newblock A note on the quadratic exponential binary distribution.
\newblock {\em Biometrika\/}~\textbf{81}(2), 403--408.

\bibitem[Danaher \emph{and others}(2014)Danaher, Wang and Witten]{danaher2014}
\textsc{Danaher, P., Wang, P. and Witten, D.~M.} (2014).
\newblock The joint graphical lasso for inverse covariance estimation across
  multiple classes.
\newblock {\em Journal of the Royal Statistical Society: Series B (Statistical
  Methodology)\/}~\textbf{76}(2), 373--397.

\bibitem[Efron \emph{and others}(2004)Efron, Hastie, Johnstone and
  Tibshirani]{Lars}
\textsc{Efron, B., Hastie, T., Johnstone, I. and Tibshirani, R.} (2004).
\newblock Least angle regression.
\newblock {\em The Annals of statistics\/}~\textbf{32}(2), 407--499.

\bibitem[El~Ghaoui \emph{and others}(2012)El~Ghaoui, Viallon and
  Rabbani]{elghaoui2012}
\textsc{El~Ghaoui, L., Viallon, V. and Rabbani, T.} (2012).
\newblock Safe feature elimination in sparse supervised learning.
\newblock {\em Pacific Journal of Optimization\/}~\textbf{8}(4), 667--698.

\bibitem[Gertheiss and Tutz(2010)Gertheiss and Tutz]{tutz2010}
\textsc{Gertheiss, J. and Tutz, G.} (2010).
\newblock Sparse modeling of categorial explanatory variables.
\newblock {\em The Annals of Applied Statistics\/}~\textbf{4}(4), 2150--2180.

\bibitem[Gertheiss and Tutz(2012)Gertheiss and Tutz]{tutz2012}
\textsc{Gertheiss, J. and Tutz, G.} (2012).
\newblock Regularization and model selection with categorial effect modifiers.
\newblock {\em Statistica Sinica\/}~\textbf{22}(3), 957--982.

\bibitem[Gross and Tibshirani(2016)Gross and Tibshirani]{gross2016}
\textsc{Gross, S. and Tibshirani, R.} (2016).
\newblock Data shared lasso: A novel tool to discover uplift.
\newblock {\em Computational Statistics and Data Analysis\/}~\textbf{101},
  226--235.

\bibitem[Guo \emph{and others}(2015)Guo, Cheng, Levina, Michailidis and
  Zhu]{Guo}
\textsc{Guo, J., Cheng, J., Levina, E., Michailidis, G. and Zhu, J.} (2015).
\newblock Estimating heterogeneous graphical models for discrete data with an
  application to roll call voting.
\newblock {\em The Annals of Applied Statistics\/}~\textbf{9}(2), 821--848.

\bibitem[Hern{\'a}n \emph{and others}(2004)Hern{\'a}n, Hern{\'a}ndez-D{\'\i}az
  and Robins]{Hernan}
\textsc{Hern{\'a}n, M.~A., Hern{\'a}ndez-D{\'\i}az, S. and Robins, J.~M.}
  (2004).
\newblock A structural approach to selection bias.
\newblock {\em Epidemiology\/}~\textbf{15}(5), 615--625.

\bibitem[H{\"o}fling and Tibshirani(2009)H{\"o}fling and
  Tibshirani]{hofling2009}
\textsc{H{\"o}fling, H. and Tibshirani, R.} (2009).
\newblock Estimation of sparse binary pairwise markov networks using
  pseudo-likelihoods.
\newblock {\em Journal of Machine Learning Research\/}~\textbf{10}(Apr),
  883--906.

\bibitem[Lauritzen(1996)Lauritzen]{lauritzen1996}
\textsc{Lauritzen, S.~L.} (1996).
\newblock {\em Graphical models\/}, Volume~17. Clarendon Press.

\bibitem[Lyons \emph{and others}(2006)Lyons, Polinder, Larsen, Mulder,
  Meerding, Toet, Van~Beeck and the Eurocost Reference~Group]{lyons2006}
\textsc{Lyons, R.~A., Polinder, S., Larsen, C.~F, Mulder, S., Meerding, W.~J.,
  Toet, H., Van~Beeck, E. and the Eurocost Reference~Group}. (2006).
\newblock Methodological issues in comparing injury incidence across countries.
\newblock {\em International journal of injury control and safety
  promotion\/}~\textbf{13}(2), 63--70.

\bibitem[Ma and Michailidis(2016)Ma and Michailidis]{ma2016}
\textsc{Ma, J. and Michailidis, G.} (2016).
\newblock Joint structural estimation of multiple graphical models.
\newblock {\em Journal of Machine Learning Research\/}~\textbf{17}(166), 1--48.

\bibitem[Meinshausen and B{\"u}hlmann(2006)Meinshausen and
  B{\"u}hlmann]{meinshausen2006}
\textsc{Meinshausen, N. and B{\"u}hlmann, P.} (2006).
\newblock High-dimensional graphs and variable selection with the lasso.
\newblock {\em The annals of statistics\/}, 1436--1462.

\bibitem[{Ollier} and {Viallon}(2014){Ollier} and {Viallon}]{Viallon_AutoArxiv}
\textsc{{Ollier}, E. and {Viallon}, V.} (2014, November).
\newblock {Joint estimation of $K$ related regression models with simple
  $L_1$-norm penalties}.
\newblock {\em ArXiv e-prints arXiv:1411.1594\/}.

\bibitem[Ollier and Viallon(2017)Ollier and Viallon]{ollier2017}
\textsc{Ollier, E. and Viallon, V.} (2017).
\newblock Regression modelling on stratified data with the lasso.
\newblock {\em Biometrika\/}~\textbf{104}(1), 83--96.

\bibitem[Tao \emph{and others}(2016)Tao, Huang, Wang, Xi and Li]{tao2016}
\textsc{Tao, Q., Huang, X., Wang, S., Xi, X. and Li, L.} (2016).
\newblock Multiple gaussian graphical estimation with jointly sparse penalty.
\newblock {\em Signal Processing\/}~\textbf{128}, 88--97.

\bibitem[Tibshirani \emph{and others}(2012)Tibshirani, Bien, Friedman, Hastie,
  Simon, Taylor and Tibshirani]{tibshirani2012}
\textsc{Tibshirani, R., Bien, J., Friedman, J., Hastie, T., Simon, N., Taylor,
  Jo. and Tibshirani, R.~J.} (2012).
\newblock Strong rules for discarding predictors in lasso-type problems.
\newblock {\em Journal of the Royal Statistical Society: Series B (Statistical
  Methodology)\/}~\textbf{74}(2), 245--266.

\bibitem[van~de Geer and B{\"u}hlmann(2013)van~de Geer and
  B{\"u}hlmann]{vandeGeer}
\textsc{van~de Geer, S. and B{\"u}hlmann, P.} (2013).
\newblock $\ell_0$-penalized maximum likelihood for sparse directed acyclic
  graphs.
\newblock {\em The Annals of Statistics\/}~\textbf{41}(2), 536--567.

\bibitem[Viallon \emph{and others}(2014)Viallon, Banerjee, Jougla, Rey and
  Coste]{viallon2014}
\textsc{Viallon, V., Banerjee, O., Jougla, E., Rey, G. and Coste, J.} (2014).
\newblock Empirical comparison study of approximate methods for structure
  selection in binary graphical models.
\newblock {\em Biometrical Journal\/}~\textbf{56}(2), 307--331.

\bibitem[Viallon \emph{and others}(2016)Viallon, Lambert-Lacroix, Hoefling and
  Picard]{ViallonFused}
\textsc{Viallon, V., Lambert-Lacroix, S., Hoefling, H. and Picard, F.} (2016).
\newblock On the robustness of the generalized fused lasso to prior
  specifications.
\newblock {\em Statistics and Computing\/}~\textbf{26}(1-2), 285--301.

\bibitem[Wainwright \emph{and others}(2007)Wainwright, Lafferty and
  Ravikumar]{wainwright2007}
\textsc{Wainwright, M.~J., Lafferty, J.~D. and Ravikumar, P.} (2007).
\newblock High-dimensional graphical model selection using $\ell_1$-regularized
  logistic regression.
\newblock In:  {\em In Advances in neural information processing systems\/}.
  pp.\  1465--1472.

\bibitem[{Wang} \emph{and others}(2009){Wang}, {Chao} and {Hsu}]{wang2010}
\textsc{{Wang}, P., {Chao}, D.~L. and {Hsu}, L.} (2009, August).
\newblock {Learning networks from high dimensional binary data: An application
  to genomic instability data}.
\newblock {\em ArXiv e-prints arXiv:0908.3882\/}.

\bibitem[Yang and Ravikumar(2011)Yang and Ravikumar]{yang2011}
\textsc{Yang, E. and Ravikumar, P.} (2011).
\newblock On the use of variational inference for learning discrete graphical
  model.
\newblock In:  {\em Proceedings of the 28th International Conference on Machine
  Learning (ICML-11)\/}. pp.\  1009--1016.

\bibitem[{Zhou} and {Zhu}(2010){Zhou} and {Zhu}]{zhou2010}
\textsc{{Zhou}, N. and {Zhu}, J.} (2010, June).
\newblock {Group Variable Selection via a Hierarchical Lasso and Its Oracle
  Property}.
\newblock {\em ArXiv e-prints arXiv:1006.2871\/}.

\end{thebibliography}

\newpage

\section{Appendix}
\begin{figure}[!ht]
	\centering
	\includegraphics[width=0.95\linewidth]{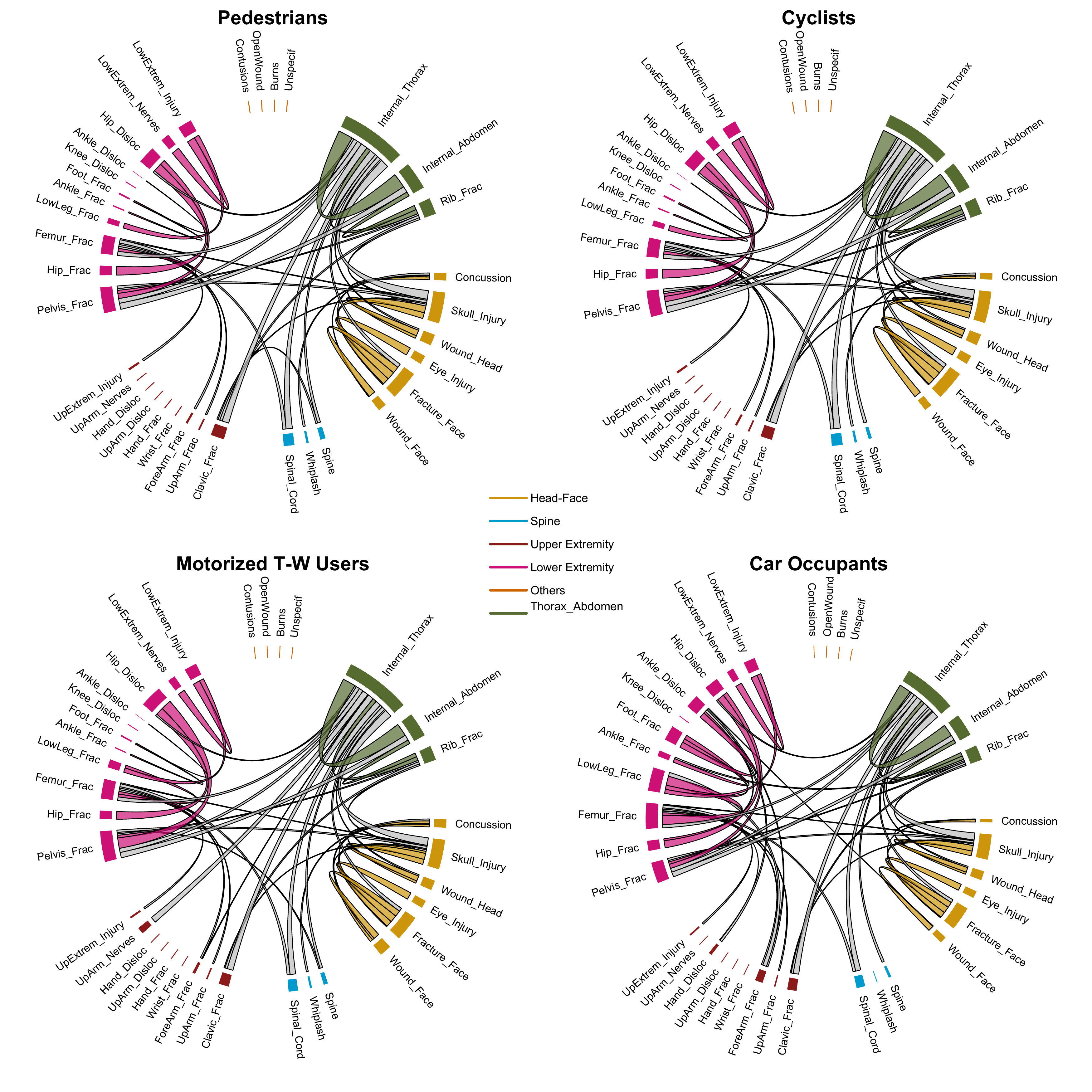}
	\caption{Application of the Data Shared SepLogit approach on the Rh\^one Registry Data to describe the injury tables of road accidents victims, according to the type of user: pedestrians, cyclists, motorized T-W and car occupants. All positive conditional associations are represented. }
	\label{App_DataSharedSepLogit_log(1)}
\end{figure}

\begin{figure}[!ht]
	\centering
	\includegraphics[width=0.95\linewidth]{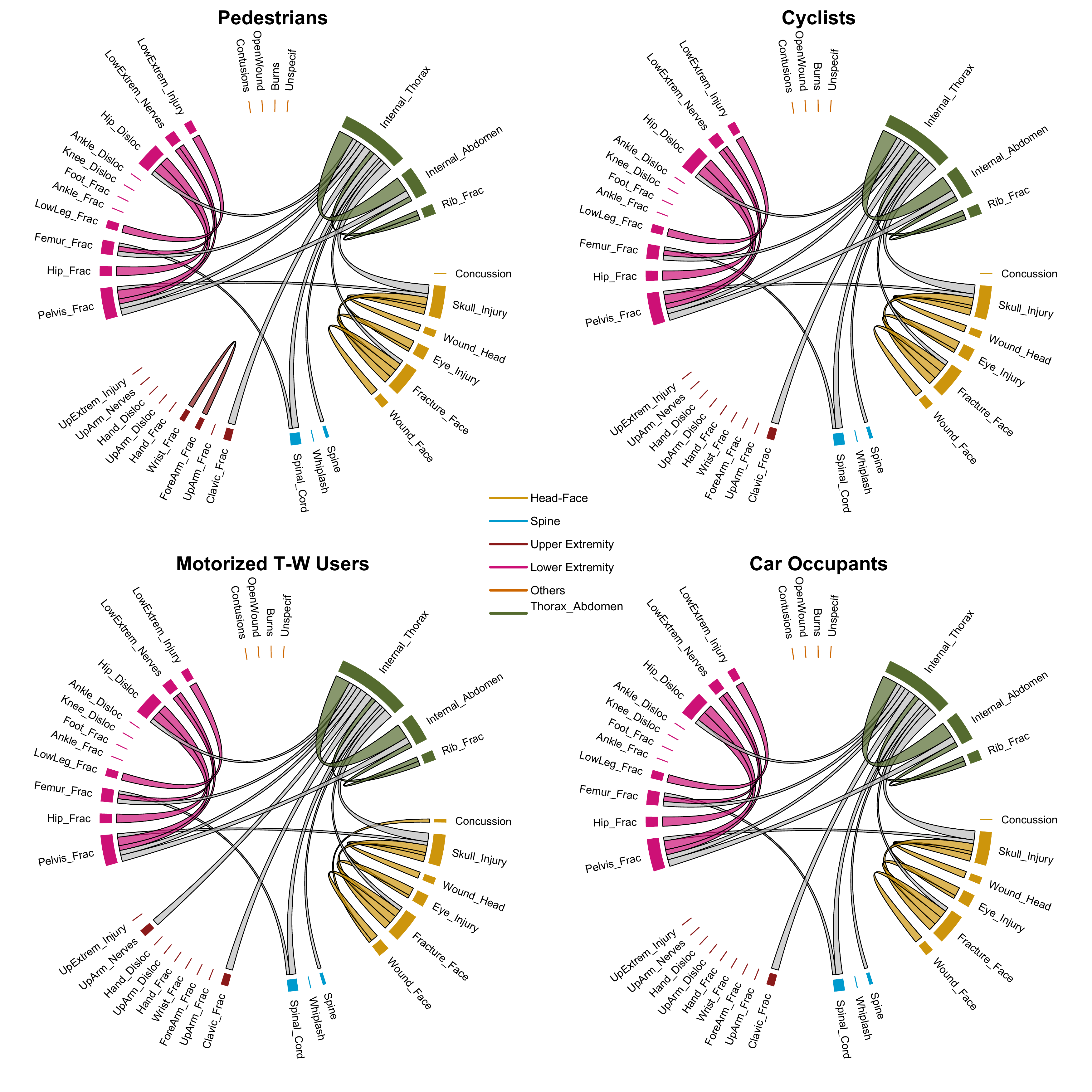}
	\caption{Application of the RefLasso SepLogit approach on the Rh\^one Registry Data to describe the injury tables of road accidents victims, according to road user type: pedestrians, cyclists, motorized T-W and car occupants. The reference stratum is that of car occupants.}
	\label{App_RefLassoSepLogit_log(2)}
\end{figure}

In Figure \ref{App_DataSharedSepLogit_log(1)}, we present another version of the structure of the graphical models estimated  by Data Shared Seplogit, on each of the four strata. Here, only positive conditional associations, that is edges corresponding to estimated conditional odds-ratios greater than one, are presented. \\
Figure \ref{App_RefLassoSepLogit_log(2)} presents the results obtained by adapting the approach of \citet{cheng2014} and applying it to the Registry data. We refer to this strategy as RefLasso SepLogit since it relies on an a priori choice for the reference stratum, that was set to car occupants here (which appears as a natural choice this application). Only edges corresponding to conditional odds-ratios greater than or equal to 2 are presented to make comparison with the results of Figure \ref{App_DataSharedSepLogit_log(2)} easier. The common structure is similar to that obtained with Data Shared SepLogit, but fewer heterogeneities are identified here. This was expected from the results of \citet*{ollier2017} since the reference stratum (car occupants) corresponds to the stratum where most heterogeneities were identified with Data Shared SepLogit: considering the model obtained by DataShared SepLogit as the true one, car occupants correspond to the worst reference stratum to work with since it leads to the highest level of heterogeneity, hence the highest complexity, and sparsistency is not guaranteed; see \citet*{ollier2017}  for more details. 

\end{document}